\documentclass[aps,pra]{revtex4}

\usepackage{latexsym}
\usepackage{amsmath}
\usepackage{graphicx}

\newcommand{\be}{\begin{equation}}
\newcommand{\ee}{\end{equation}}
\newcommand{\Tr}{ {}{\rm Tr}{} }
\def\JPA{{J.\ Phys.\ A\ }}
\def\ADP{{Adv.\ Phys.\ }}
\def\JPC{{J.\ Phys.\ C\ }}
\def\NPB{{Nucl.\ Phys.\ B\ }}
\def\PRB{{Phys.\ Rev.\ B\ }}
\def\PRL{{Phys.\ Rev.\ Lett.\ }}

\def\ZPB{{Z.\ Phys.\ B\ }}

\begin{document}

\title{One-step replica symmetry breaking solution of the quadrupolar glass
model}

\author{F. P. Mancini}
\address{Dipartimento di Fisica and
 Sezione I.N.F.N., Universit\`a di
Perugia, Via A. Pascoli, Perugia, 06123, Italy}
\author{D. Sherrington}
\address{Rudolf Peierls Centre for Theoretical Physics, University
of Oxford,1 Keble Road, Oxford, OX1 3NP, UK}

\begin{abstract}
We consider the quadrupolar glass model with infinite-range random
interaction. Introducing a simple one-step replica symmetry
breaking ansatz we investigate the para-glass continuous
(discontinuous) transition which occurs below (above) a critical
value of the quadrupole dimension $m^*$. By using a mean-field
approximation we study the stability of the one-step replica
symmetry breaking solution and show that for $m>m^{*}$ there are
two transitions. The thermodynamic transition at a temperature
$T_{D}$ is discontinuous but there is no latent heat. At a higher
temperature we find the dynamical or glass transition temperature
$T_G$ and the corresponding discontinuous jump $q_G$ of the order
parameter.
\end{abstract}


\maketitle

\section{INTRODUCTION}
In the last decades quadrupolar glasses have found widespread
experimental and theoretical interest \cite{hochli_90}. Disordered
quadrupolar glasses are produced by random dilution of
(quadrupolar) molecular crystals with atoms which have no
quadrupole moments; well-known examples of such systems are
$K(CN)_{x}Br_{1-x}$ or $Na(CN)_{x}Cl_{1-x}$, or $N_{2} \: Ar$,
$CuCN$, or solid hydrogen (see Ref. \cite{binder_92} for a
review). The success of the Sherrington-Kirkpatrick (SK) model
\cite{sk} in providing a good theory to describe systems of
interacting magnetic or electric dipole moments, suggests to
extend to quadrupolar glasses the same kind of analysis. However,
there are differences between standard spin glass systems and
quadrupolar glasses; the latter do not have the global inversion
symmetry $S_i \to - S_i$ for all spins. For several systems
without reflection symmetry, and close to the transition
temperature, only one step in the Parisi replica symmetry breaking
scheme is sufficient to describe the transition para-glass above a
lower critical dimension \cite{gold_elder_85}. Indeed, the
one-step replica symmetry breaking ($1RSB$) scheme has proven to
provide stable solutions for the Potts glass model
\cite{gks,ck,desantis}, the spherical $p$-spin model
\cite{crisanti_92} and the $p$-spin Ising spin glass model
\cite{gardner_85}. It is the purpose of the present paper to show
that the $1RSB$ scheme can be applied also to the quadrupolar
glass model, and indeed it provides a stable solution in certain
regimes.


In the present paper we shall consider a perturbative evaluation
of the free energy by means of a Taylor expansion up to fourth
order in the order parameter. It is obvious that the perturbative
approach is most reliable near the transition temperature. We
shall show that the transition from the replica symmetric ($RS$)
state to the $1RSB$ occurs either discontinuously or continuously,
depending on the value of the quadrupole dimension $m$. A similar
dependence of the $RS$ to $1RSB$ transition, though on the value
of an external field, is exhibited by the spherical $p$-spin model
\cite{crisanti_92} and  the Ising $p$-spin model
\cite{gardner_85}. For any $p>2$, the transition is discontinuous
(continuous) for fields weaker (larger) than a critical value of
the external field $h_c$, which depends on $p$.

The plan of the paper is as follows: in Sec. II
we shall use a rather pedagogical approach mainly to review the results obtained
in the mean-field analysis of the quadrupolar glass model in the
framework of the replica symmetry ansatz \cite{gs}. Section III is devoted to
the study of the $1RSB$ solutions of the
saddle-point equations, assuming all the transitions to be continuous or at worst
weakly discontinuous. In Sec. IV we shall perform the de Almeida-Thouless (AT)
stability analysis while the dynamical transition is discussed in Sec. V. The
concluding remarks are given in Sec. VI.


\section{Uniaxial Quadrupolar Glass}\label{section_UQG}
\noindent
The infinite-ranged quadrupolar glass model has been first introduced by
Goldbart and Sherrington (GS) \cite{gs}. The model assumes the quadrupole-quadrupole
interaction to be more dominant than the interactions between dipoles.
This appears to be the case in several experimental situations, where the
quadrupolar species occupy the sites of a regular lattice, but share this lattice with a
dilutant without quadrupole moment: argon in the case of interacting $N_{2}$, parahydrogen
in the case of interaction with orthohydrogen, and $K Br$ in the case of $KCN$, etc.
\cite{binder_92}.

To construct the mean-field theory of a set
of uniaxial quadrupoles interacting through randomly quenched and
frustrated isotropic exchange one may adopt the Hamiltonian
\begin{equation}
H = -\sum_{(i,j)} J_{ij} \sum_{\mu \nu} S_{\mu}^i S_{\nu}^i S_{\mu}^j
S_{\nu}^j = - \sum_{(i,j)} J_{ij} \left( \mbox{\bf S}^i \cdot \mbox{\bf S}^j
\right)^2  ,
\label{quadrup_ham}
\end{equation}
where the spin vector ${\bf S}^i$ is defined via the component $f^{i}_{\mu
\nu}= \left(S^{i}_{\mu}S^{i}_{\nu}-\delta_{\mu \nu}/m \right)$ of the electric
quadrupole moment tensor \cite{gs}. The summation $(i,j)$ runs over
all the distinct pairs. Each ${\bf {S}}^i$ has $m$ components $S^i_{\mu}$ ($\mu
=1, \ldots ,m$) and, for convenience, is assumed to be a vector with fixed
length $\vert {\bf {S}}^i \vert=m$. Of course, taking general $m$ does not
describe the experimental quadrupolar glasses. Rather it is a natural model to
consider theoretically for classification. By analogy with SK, the spins are taken
to interact via independent random interactions $J_{ij}$ which are assumed
Gaussian distributed:
\begin{equation}
P \left( J_{ij} \right) = \left( \frac{N}{2 \pi J^2} \right)^{1/2} \exp
\left[ - \frac{(J_{ij}- J_0 /N)^2}{2 J^{2}/N} \right] .
\end{equation}
The mean $J_0$ and the variance $J$ of the
distribution depend on the total number of quadrupoles $N$ to ensure a
meaningful thermodynamic limit $(N \rightarrow \infty)$ with an extensive
energy: $J_0 = \tilde{J}_0/N$ and $J=\tilde{J}/N^{1/2}$.
The Hamiltonian (\ref{quadrup_ham}) can be seen as the Hamiltonian of an
infinite-range model for $N$ classical vector spins ${\bf {S}}^i$ and
zero external field is assumed.

In Ref. \cite{gs} it has been shown that in terms of the order parameters
\be
\begin{split}
Q^{ab}_{\mu \nu \lambda \rho} & =  \frac{ \mbox{tr } \left[ S_{\mu}^{a}
S_{\nu}^{a} S_{\lambda}^{b} S_{\rho}^{b} \: \exp L \right]} { \mbox{tr }
\left[ \exp L \right]} = \langle S_{\mu}^{a} S_{\nu}^{a} S_{\lambda}^{b}
S_{\rho}^{b} \rangle \\
M^{a}_{\mu \nu} &=  \frac{ \mbox{tr } \left[S_{\mu}^{a} S_{\nu}^{a}\: \exp
L \right] } { \mbox{tr } \left[ \exp L \right] } = \langle S_{\mu}^{a}
S_{\nu}^{a} \rangle ,
\label{tensors}
\end{split}
\ee
the free energy per spin $f$ - by means of the replica trick - is given by
\be
\begin{split}
- \beta f &= \frac{1}{N} \lim_{n \rightarrow 0} \frac{1}{n} \left(
\overline{Z^n} - 1 \right)
 = \lim_{n \rightarrow 0} \frac{1}{n} \left[ -\frac{\beta J_0}{2}
\sum_{a} \sum_{\mu \nu} \left( M^{a}_{\mu \nu} \right)^2 \right.
\\
&- \left. \left( \frac{\beta J}{2} \right)^2 \sum_{ab} \sum_{\mu \nu
\lambda \rho} \left( Q^{ab}_{\mu \nu \lambda \rho} \right)^2 + \mbox{log
tr } \exp L \right],
\end{split}
\label{fee}
\ee
where $n$ is the number of replicas, $\beta=1/k_{B}T$, and $L$ is
\begin{equation}
L= \beta J_0 \sum_{a} \sum_{\mu \nu} M^{a}_{\mu \nu} S^{a}_{\mu}S^{a}_{\nu}+
\frac{ ( \beta J)^2}{2} \sum_{ab} \sum_{\mu \nu \lambda \rho} S_{\mu}^{a}
S_{\nu}^{a} S_{\lambda}^{b} S_{\rho}^{b} Q^{ab}_{\mu \nu \lambda\rho} .
\label{elle}
\end{equation}
The elements of the order parameters $Q^{ab}_{\mu \nu \lambda
\rho}$ and $M^{a}_{\mu \nu}$ are not independent quantities and
they can be parameterized in terms of five sets of independent
parameters $A^a$, $B^{ab}$, $C^{ab}$, $D^{ab}$, and $E^{ab}$. The
non-zero extremal values of the above sets describe possible glass
ordering \cite{gs}.

Upon decreasing the temperature, when one of the parameters becomes
different from zero, the high-temperature disordered phase becomes
unstable. By assuming continuous transition in the
replica symmetric ansatz and provided that the
average interaction is not too positive,
\begin{equation}
 \frac{J_{0}}{J}<
\frac{-m^{2}+m+8}{m+4},
\label{ferro}
\end{equation}
there is a transition to an isotropic glass state occurring at the temperature $T_{RS}=(2m
J)/[k_{B}(m+2)]$.  This highest temperature phase transition is associated with the order
parameter $B^{ab}$ acquiring a non-zero value, hinting at the presence of
isotropic quadrupolar order \cite{gs}.
As GS showed, below $T_{RS}$ the
replica symmetric solution isotropic glass phase is unstable with respect to
fluctuations  in the space of broken replica symmetric isotropic
glass order parameters, the instability being stronger than in conventional
systems.

It is worth noting that Eq. (\ref{ferro}) is the equivalent of the
condition $J_{0}/J<1$ in the {\it SK} model to ensure a transition from
the paramagnet state to spin glass state. Furthermore, when the
numerator of the right hand side of Eq. (\ref{ferro}) becomes negative,
it is necessary to introduce a negative value for $J_{0}$ in order to
obtain a glass phase at low temperatures; this happens for $m>3.37
\ldots$. Increasing the negativity of $J_{0}$ reduces the
temperature of ferromagnetism onset, though it cannot be stopped.


\section{One-step replica symmetry breaking Theory}
\noindent
The further analysis of GS leads to the conclusion that a replica
symmetric ansatz cannot give a stable, and hence physical, solution of
the quadrupolar glass model. Thus, one has to resort to a
replica symmetry breaking ansatz.
Close to the isotropic quadrupolar glass transition, i.e. confining
attention to regions of parameter space $(J_0, J)$ in which the highest
transition temperature does correspond to a phase transition in the
order parameter $B^{ab}$, it is sufficient to consider only this
parameter different from zero in the free energy (see Ref.
\cite{gs}). Thus, one requires $J_0$ to satisfy the inequality
(\ref{ferro}) for $T \lesssim T_{RS}$, i.e. in the neighbourhood of the
transition temperature. The free energy is then given by
\begin{equation}
\begin{split}
\beta f &=
 \lim_{n \rightarrow 0}
\frac{1}{n}\left\{ \frac{\left( \beta J\right)
^{2}}{2}(m-1)(m+2){\sum_{ab}}^{\prime} \left( B^{ab}\right)^2 +m\left( \beta
J\right)^2 {\sum_{ab}}^{\prime}B^{ab}\right.
\\
&- \left.\log \Tr \: \exp
\left[ \left( \beta J\right)^2 {\sum_{ab}} ^{\prime} \sum_{\mu \nu}
B^{ab}S^{a}_{\mu} S^{a}_{\nu}S^{b}_{\mu} S^{b}_{\nu} \right]\right\} .
\end{split}
\label{fe_sa}
\ee
The symbol $\sum_{ab}^{\prime }$ stands for a sum which excludes terms
with any equal indices; i.e. $a \neq b$. The paramagnetic
contribution $\beta f_{PM}=-(\beta \: J)^2 m(m-3)/2$ has been subtracted as it does not
depend on the order parameter. One then looks for a
replica symmetry breaking ansatz for $B^{ab}$.  Here we consider this to the first
level in the standard Parisi procedure.

Using the standard procedure of the replica symmetry breaking
method, one groups the $n$ replicas in blocks of $x$, where $x$ is
a parameter (between $1$ and $n$) to be located by the saddle
points equations. Each block contains $x$ replicas. Thus, one has
\begin{equation}
B^{ab}=
\begin{cases}
q & \text{if $I(a/x)=I(b/x)$} \\
0 & \text{otherwise}
\end{cases}
\label{rsba}
\end{equation}
where $I(y)$ is an integer valued function: its value is the smallest
integer which is greater than or equal to $y$.
Upon substituting in Eq. (\ref{fe_sa}), one has
\begin{equation}
\begin{split}
\beta f &= \frac{\left( \beta J\right) ^{2}}{2}(m-1)(m+2)(x-1)q^2 +\left(
\beta J\right) ^{2}(x-1)m q +\left( \beta J\right) ^{2}m^2 q
\\
&- \lim_{n \rightarrow 0} \frac{1}{n}\log \Tr \exp \left[ \left( \beta
J\right)^2 q \sum_{\mu \nu} \sum_{k}^{n/x} \left( \sum_{a \subset
block(k)}S^{a}_{\mu} S^{a}_{\nu} \right)^2 \right] ,
\end{split}
\label{fe_rsb}
\ee
where there are $n/x$ blocks labelled by $k=1, \ldots, n/x$. The
index $a$ belongs to block $k$ if: $I(a/x)=k$.
We shall now perturbatively compute the free energy (\ref{fe_rsb}) by
Taylor-expanding the free energy up to fourth order in $q$. This
approximation implies that  one is assuming the transitions to be
continuous or at worst weakly discontinuous.
After some lengthy but straightforward algebra,
one obtains the following expression for the free energy:
\begin{equation}
\begin{split}
\beta f (q,x) &=-\frac{\alpha_{2}}{2}\:(x-1)\:t \:
q^{2}-\frac{\overline{\alpha}_{3}}{3}\:(x-1) \:
q^{3}-\frac{\hat{\alpha}_{3}}{3}\:(x-1)(x-2) \: q^{3} \\
&-\frac{\beta_{1}}{12}\:(x-1)q^{4}-\frac{\beta_{3}}{12}\:(x-1)(x-2)q^{4}-
\frac{\beta_{5}}{12}\:(x-1)(x-2)(x-3)q^{4}+ O[q^5],
\end{split}
\label{functional}
\end{equation}
where the coefficients in the free energy expansion are given by
\be
\begin{split}
\alpha_2 &= \frac{(m-1)(m+2)^3}{4m^2 \left(1-t\right)^2}  \\
\overline{\alpha}_3 &=\frac{(m-1)(m-2)(m+2)^5}
{4m^3(m+4)\left(1-t\right)^3} \\
\hat{\alpha}_3 &= \frac{(m-1)(m+2)^4}{4m^3\left(1-t\right)^3} \\
\beta_1 &= \frac{3(m-1)(m-4)(m+2)^6(m^2 +m -3)}
{4m^4(m+4)(m+6)\left(1-t\right)^4} \\
\beta_3 &= \frac{3(m-1)(m-2)(m+2)^6}{m^4(m+4)\left(1-t\right)^4} \\
\beta_5 &= \frac{3(m-1)(m+2)^5}{4m^4\left(1-t\right)^4} ,
\end{split}
\ee and $t$ is the reduced temperature, defined as
$t=1-\left(T^2/T_{C}^2\right)$, where
\be
 T_{C}=\frac{2m
\,J}{(m+2)k_B} .
\ee
For $m$ less than a critical value $m^*$, discussed below, the transition is
continuous, with the same $T_{C}$ as predicted by $RS$ theory \cite{gs}. In this case
one may use the saddle point method to evaluate the extremal values of the
parameters $q$ and $x$. These saddle point equations are
\begin{equation}
\frac{\partial f}{\partial q}=\frac{\partial f}{\partial x}=0,
\nonumber
\end{equation}
from which one has
\be
\begin{split} 0&= 3t\alpha _{2}+3q(\overline{\alpha}_{3}+(x-2)
\hat{\alpha } _{3})+q^{2}\left[ \beta _{1}+(x-2)(\beta_{3}+(x-3)
\beta _{5}) \right]
\\
0 &= 6t\alpha _{2}+4q(\overline{\alpha}_{3}+(2x-3)
\hat{\alpha } _{3})+q^{2}\big[\beta _{1}+(2x-3)\beta_{3}
+(3x^{2}-12x+11)\beta _{5}\big].
\end{split}
\label{sad_point}
\ee
According to the value of the quadrupole dimension $m$, the transition from the
higher temperature $RS$ phase can occur either continuously or discontinuously.
The transition is continuous in $q$ for $m<m^*\simeq 3.37$, i.e. when the coefficient
$\hat{\alpha}_{3}$ becomes larger than $ \overline{\alpha}_{3}$.  $q$ and $x$ satisfy the
following equations that  express the extremum of the free energy functional
(\ref{functional})
\begin{subequations}
\begin{eqnarray}
t<0 &:&
\begin{cases}
q  &=0 \\
x  & \text{undetermined}
\end{cases}
\label{s_p_sol_1}
\\
t>0 &:&
\begin{cases}
q & =-\frac{t\alpha _{2}}{2(\overline{\alpha }_{3}-\hat{\alpha }_{3})}
+
\frac{t^{2}\alpha _{2}}{48\left(\overline{\alpha }_{3}-\hat{\alpha}_{3}\right)^{3}\:
 \hat{\alpha}_{3}^{2}}
\left\{ \left( \overline{\alpha }_{3}^{2}+4\overline{
\alpha }_{3}\hat{\alpha }_{3}-15\hat{\alpha }_{3}^{2}\right) \beta _{5}
\right.
\\
& + \left. \hat{
\alpha }_{3}\left[ -5\beta _{1}\hat{\alpha }_{3}+\beta _{3}\left( 7\hat{
\alpha }_{3}-2\overline{\alpha}_{3}\right) \right] \right\}
 +O \left[ t^{3}\right],
\\
x & =\frac{\overline{\alpha }_{3}}{\hat{\alpha }_{3}}+\frac{t\alpha _{2}
\left[ \hat{\alpha }_{3}^{2}(-\beta _{1}+\beta _{3})+\beta _{5}(\overline{
\alpha }_{3}^{2}-2\overline{\alpha }_{3}\hat{\alpha }_{3}-\hat{\alpha }
_{3}^{2})\right] }{4(\overline{\alpha }_{3}-\hat{\alpha }_{3})\hat{\alpha }
_{3}^{3}}+O[t^{2}].
\end{cases}
\label{s_p_sol_2}
\end{eqnarray}
\label{s_p_sol_tutti}
\end{subequations}
Close to the transition temperature $T_C$, i.e. when  $t\gtrsim 0$, the
solution(\ref{s_p_sol_2}) is valid only for $2<m\le m^{*}\simeq 3.37$ within $1RSB$
subspace. At $m^*$ the cubic term in the free energy functional
(\ref{functional}) changes sign. This coincides with $x=1$.  Thus, above
$m^*$, the transition is not anymore continuous. The parameters  $q$ and
$x$ are plotted for $m=3$ within this approximation in Fig.
\ref{quadru_q_x_T}.
\begin{figure}
\begin{center}
\includegraphics[scale=.35]{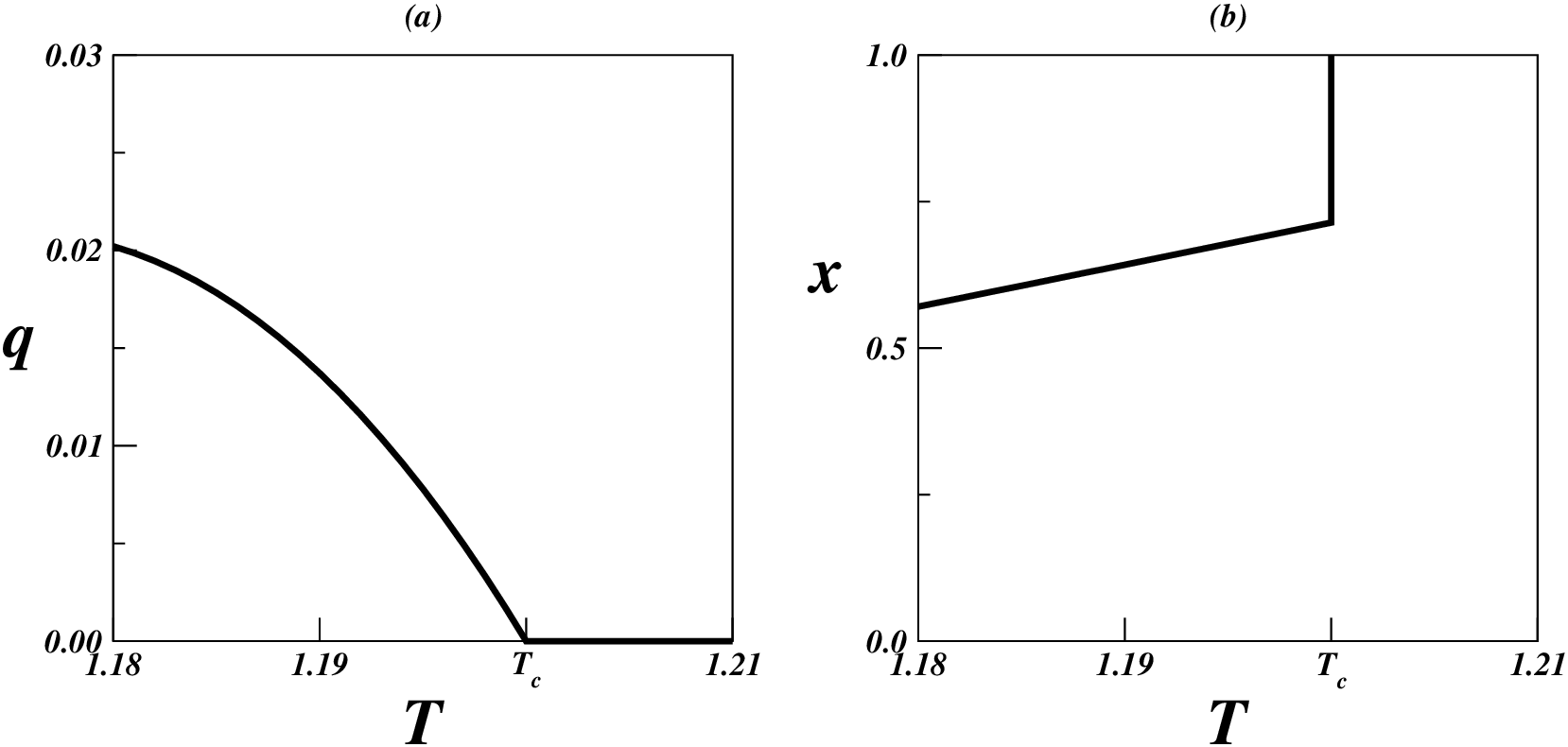}
\caption{\label{quadru_q_x_T} (a) Plot of the one-step parameter $q$ as
a function of the temperature $T$ (in units of $k_B/J$); (b) Plot  of the one-step breaking
parameter $x$  as a function of the temperature $T$ (in units of $k_B/J$).
Both plots are drawn for $m=3$ and within the quartic approximation for
the free energy (\ref{functional}).}
\end{center}
\vspace{8mm}
\end{figure}
By substituting Eq. (\ref{s_p_sol_2}) in Eq. (\ref{functional}) one finds
the free energy of the glass phase close to the transition, to be
\be
\beta f = \frac{\alpha_2^3 \: t^3}{24 \hat{\alpha }_{3}\left(\hat{\alpha
}_{3}- \overline{\alpha }_{3}\right)} + O[t^{4}]=
-\frac{(m-1)(m+2)(m+4)\: t^3}{96(m^2-m-8)} + O[t^{4}]  .
\ee
Below the transition the free energy is larger than that of
the paramagnetic phase.

On the other hand, when $m>m^{*}$ there is still a glass solution to the saddle
point equations but with a discontinuous onset of $q$ from the higher
temperature $RS$ phase. The
transition may be found with the additional requirement that
the free energy in the paramagnetic phase is equal to the one in
the glass phase with break point $x$ equal to $1$. Denoting this
transition temperature by $T_{D}$, to the quartic order in $q$
for the free energy, one finds that
\be
t_{D}=\frac{2(\overline{\alpha }_{3}-\hat{\alpha}_{3})^{2}}{3\alpha _{2} (\beta _{1}-\beta
_{3}+2\beta _{5})}=\frac{2(m+6)(m^2-m-8)^2}{9(m+4)(168+ 34 m - 35 m^2 - 5 m^3 +m^4)} ,
\label{t_disc}
\ee
where $t_{D}=1-\left(T_{D}/T_{C}\right)^2$. There is a discontinuous
jump in the order parameter at $T_{D}$ from zero to
\be
q_{D}=-\frac{2(\overline{\alpha }_{3}-\hat{\alpha }_{3})}{\beta
_{1}-\beta _{3}+2\beta _{5}}
=
-\frac{2m(m+6)(m^2-m-8)(1-t_{D})}
{3(m+2)(168+ 34 m - 35 m^2 - 5 m^3 +m^4)}.
\label{q_disc}
\ee
where $q_{D}=q(T=T_{D}^-)$. In the neighbourhood of the
transition temperature $T_{D}$ one finds that $(x-1) \propto
(t-t_{D})$ and that the free energy of the quadrupolar glass phase is
given by
\be
\beta f=q_{D} \left(t-t_{D}\right)^{2}+O\left[
\left(t-t_{D}\right)^{3} \right] .
\nonumber
\ee
Even though the transition is discontinuous in the order parameter $q$, there
is no discontinuity in any thermodynamic quantities. Moreover, there is no
latent heat at the transition. This behaviour is qualitatively common to a whole
class of mean-field models of spin glasses e.g. the $p>2$ spin model beneath a
critical field \cite{gardner_85} and the Potts glass model above the critical
Potts dimension $p=4$ \cite{gks,kirkwoly_87,ck}.

Since the order parameter has a discontinuous
jump at the transition temperature  when $m>m^{*}$,
the perturbative approach should not be valid anymore.
However, one may control the approximation by setting $
m=m^{*}+\varepsilon$, where $\varepsilon \ll 1$. Thus, one obtains a quadrupolar
glass phase with broken replica symmetry appearing below $t_{D} \propto
{\varepsilon}^{2}$, with $q_{D} \propto \varepsilon$ and $x(T \to
T_{D}^{-}) \to 1$. Explicitly, to leading order, one has
\begin{subequations}
\be
q_{D}(m=m^*+\varepsilon) =\frac{2m^{*} g(m^*)  (m^{*}+4)(1-2m^{*}) }
{m^{*}+2}\,\varepsilon +O[\varepsilon^{2}],
\label{q_RSB}
\ee
\be
t_{D}(m=m^*+\varepsilon) =4 g(m^*)(m^{*2}-m^{*}-5/2)
\,\varepsilon^{2}+O[\varepsilon ^{3}],
\label{t_RSB}
\ee
\end{subequations}
where
$g(m^*)=(m^*+6)/\left[3(m^*+4)(m^{*4}-5m^{*3}-35m^{*2}+34m^{*}+168)\right]<0$. In
the next section, we shall investigate the stability of the $1RSB$ solution found
 against small further $RSB$ fluctuations.


\section{Stability analysis}
\noindent In order to study the stability of the $1RSB$ ansatz one
introduces $1RSB$-breaking fluctuations
\be
 B^{ab}=q \, \delta_{G_{a}G_{b}}+\eta ^{ab} ,
\label{fluct}
 \ee
and expands the free energy to second order in the fluctuations
$\eta^{ab}$ \cite{at}. The group Kronecker delta $
\delta_{G_{a}G_{b}}$ is unity if $a$ and $b$ belong to the same
group and zero otherwise \cite{ck}.  One has to compute the second
derivatives of the free energy (\ref{fe_sa}) with respect to
$\{B^{ab}\}$ at the $1RSB$ solution. The Taylor expansion of the
free energy around the $1RSB$ solution is
\be
f=f(q_D\, \delta_{G_{a}G_{b}})+\frac{\partial f}{\partial B^{ab}}
\; \eta^{ab}+ \frac{\partial^2 f}{\partial B^{ab}\partial
B^{cd}} \; \eta^{ab} \eta^{cd}+...
\ee
The quadratic form
\be
\Delta =\frac{\partial^{2}f}{\partial B^{ab}\partial B^{cd}}
\, \eta^{ab}\eta^{cd}
\label{Delta}
\ee
should be positive definite for a stable solution of the problem. It is easy to
see that the only non-vanishing terms are the ones where $a$, $b$, $c$ and $d$
all belong to the same group and the ones where $a$ and $c$ belong to group $k$,
and  $b$ and $d$ belong to group $k'$. The
Hessian matrix $S$ associated with the quadratic form (\ref{Delta}) factorizes
in $n/x$ identical submatrices of dimension $x(x-1)/2\times x(x-1)/2$ which
couple intragroup fluctuations and $(n/x) \: \left[ (n/x) -1\right]/2$
identical submatrices of dimension $x^{2}\times x^{2}$ which couple
intergroup fluctuations. We shall give details only on the intragroup
matrices, because lengthy computations show that the intergroup matrices
always have positive eigenvalues in the range of validity of the
solution (see also Ref. \cite{ck}).

If the $1RSB$ is stable thermodynamically, all of the eigenvalues of the stability matrix
should be positive. The intragroup matrices $S^{(ab)(cd)}$ have three different types of
matrix element:
\begin{eqnarray}
S^{(ab)(ab)} &=&-2\alpha
_{2}t-4\overline{\alpha }_{3}B^{ab}-2\beta _{1}(B^{ab})^{2}-\frac{1}{3}\beta_3
\sum_{k\neq a\neq b}B^{bk}B^{ka} \nonumber
\\
S^{(ab)(ac)} &=&-2\hat{\alpha }_{3}B^{bc}-\frac{1}{6}\beta _{3} \left[
(B^{bc})^2 +2 B^{ab} B^{bc}+2 B^{bc}B^{ca} \right] \nonumber
\\
&-& \frac{2}{3}\beta_5
\sum_{k\neq a\neq b \neq c} B^{kb}B^{ck} \nonumber
\\
S^{(ab)(cd)} &=& - \frac{2}{3} \beta_5 \left( B^{bc} B^{da}+ B^{ac}B^{db}
\right),  \nonumber
\end{eqnarray}
where $a\neq b\neq c\neq d$. Since $a$, $b$, $c$ and $d$ belong to same
group
\begin{eqnarray}
S^{(ab)(ab)} &=&-2\alpha _{2}t-4\overline{\alpha }_{3}q- 2\beta _{1}q^{2}-
\frac{1}{3}\beta_3 (x-2)q^2  \nonumber
\\
S^{(ab)(ac)} &=& -2\hat{\alpha}_{3}q-\frac{5}{6}\beta_{3} q^2 -\frac{2}{3}
\beta_5 (x-3) q^2   \nonumber
\\
S^{(ab)(cd)} &=& - \frac{4}{3} \beta_5 q^2 .  \nonumber
\end{eqnarray}
The eigenvalues of the intragroup matrices, to order $q^2$, are given
by
\be
\begin{split}
\lambda_{1} &=-2t{{\alpha}_2} -   4q \left[ \overline{\alpha }_{3}+
(x-2)\hat{\alpha}_3 \right] -
2q^2\big[\beta_1 + (x -2){{\beta}_3} +  (x^2-5x+6){{\beta}_5} \big]
\\
\lambda_{2} &=-2t \,\alpha_2  - 2q \left[2\overline{\alpha }_{3} + (x
-4){\hat{\alpha}_3} \right]  - \frac{1}{6}\, q^2\left[12{{\beta}_1}
-\big(24 - 7x \big)\beta_3 + 4\big(x^2-9x+18\big)\beta_5
\right] \\
\lambda_{3} &=-2 t \alpha_2 +   4q\big(  \hat{\alpha}_{3}
-\overline{\alpha }_{3}\big) - \frac{1}{3}\, q^2 \, \left[ 6{{\beta }_1}
+ \big( x -7\big) {{\beta }_3} - 4 (x -4) {{\beta }_5} \right].
\end{split}
\label{eigenvalues}
\ee
The behaviour of the eigenvalues in the ordered phase is obtained
by substituting  in to the above equations the values of the
parameters $q$ and $x$ pertinent to the continuous or
discontinuous transition. Close to the continuous transition
temperature, one finds that the first two eigenvalues are positive
in the range of validity of the solution $2<m<m^{*}$:
\be
\begin{split}
\lambda_1 &= \frac{(m-1)  (m+2)^3}{2 m^2} \, t +O[t^2]
\\
\lambda_2 &=\frac{(m+2)^3 \left(m^3-3 m^2-10 m+12\right)}{4 m^2 \left(m^2-m-8\right)} \, t
+O[t^2]
.\end{split}
\ee
The last one, to order $t^2$,
\be
\lambda _{3}= \frac{(m-1) (m+2)^3 \left(2m^5 + 15m^4 - 8m^3- 104m^2- 32m
+96 \right)}{16m^2(m+6)(m^2 -m-8)^2}\, t^2  +O[t^3]  ,
\ee
is positive only for $m>m_{2}^{*}\simeq 2.46$. Thus, one finds a $1RSB$ stable mean-field
theory with a continuous transition only in the range $m_{2}^{*}<m<m^{*}$. This lower limit
is the same as given in Ref. \cite{gold_elder_85} and obtained by means of  a complementary
calculation based on the full replica symmetry breaking ($FRSB$) ansatz near $T_C$ with a
perturbation treatment. For $m>m^*$ the behaviour of the eigenvalues in the ordered phase is
obtained by setting $x=1$ and by substituting $q$ with the value $q_D$  obtained in Eq.
(\ref{q_RSB}). Upon inserting these values, one easily finds that all the fluctuations
around the ordered $1RSB$ phase are finite and positive:
\be
\begin{split}
\lambda_1 &=\lambda_3 =
\frac{g(m^*)  \left(m^*-1\right) \left(m^*+2\right)^3
    \left(-10 m^{*2}+10 m^* +3\right)}{m^{*2} }
   \: \varepsilon^2  >0
\\
 \lambda_2 &=\frac{ g(m^*) \left(m^*-4\right) \left(m^*-1\right)
  \left(m^*+2\right)^3  \left(2 m^*-1\right) \left(2
 m^*+5\right)}{ m^{*2} } \: \varepsilon>0 .
 \end{split}
\ee
Thus, within our approximation, one finds a $1RSB$ stable
mean-field theory with a discontinuous transition when  $m>m^*$.


\section{Dynamical transition}
Generally, disordered systems with a discontinuous transition have a
temperature $T_G$ where a dynamic instability appears. This temperature is
called the glass temperature and is higher than the transition
temperature $T_{D}$ where the replica symmetry breaks thermodynamically, if the
latter breaking is discontinuous.

In the soft spin version of the Potts glass model \cite{kirkthiru}
it has been shown - by means of dynamical studies of the
mean-field theory - that indeed there is another transition at
temperature $ T_{G}>T_{D}$ as in the $p$-spin model for $p>2$
\cite{kirkthiru2}. Both static and dynamic transitions in the
Potts ($p>4)$ case, have also been found in Refs.
\cite{ck,desantis,kirkwoly_87}. In the study of the thermodynamics
of the quadrupolar glass, $T_G$ can be computed by means of
marginal stability \cite{marg_stabili}. By requiring the vanishing
of the first and second derivative of the free energy
(\ref{functional}) with respect to $q$ \be \left( \frac{\partial
f}{\partial q} \right)_{q=q_G} = 0 \quad  \quad ;\quad \quad
\left( \frac{\partial^2 f}{\partial q^2} \right)_{q=q_G} = 0 .
\label{marg_stab_eqs} \ee one finds, within our approximation, the
dynamical transition temperature $T_G$ and the corresponding
discontinuous jump $q_G$ of the quadrupolar glass model \be t_G =
\frac{3(\overline{\alpha}_3- \hat{\alpha}_3)^2}{4\alpha_2 (
\beta_1- \beta_3 +2\beta_5)} = \frac{(m+6)(m^2-m-8)^2}
{4(m+4)(168+ 34 m - 35 m^2 - 5 m^3 +m^4)} \label{t_G} \ee and \be
q_G = -\frac{3(\overline{\alpha}_3-
\hat{\alpha}_3)}{2(\beta_1-\beta_3 +2\beta_5)}
 = -\frac{m(m+6)(m^2-m-8)(1-t_{G})}
{2(m+2)(168+ 34 m - 35 m^2 - 5 m^3 +m^4)} ,
\ee
where $t_{G}=1-\left(T_{G}/T_{C}\right)^2$ and $q_{G}=q(T=T_{G}^-)$.
\begin{figure}
\begin{center}
\includegraphics[scale=.35]{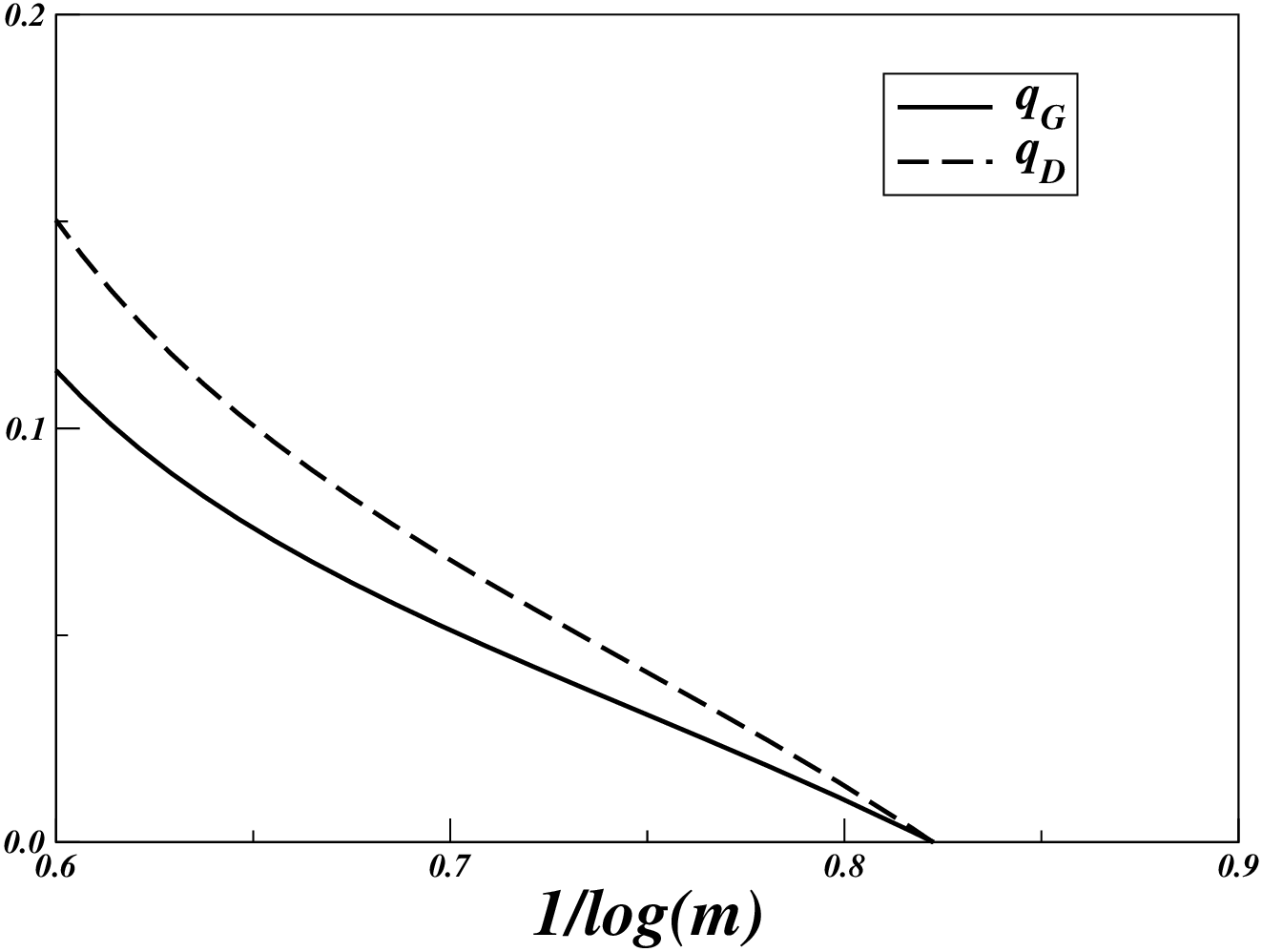}
\caption{\label{stat_dyn_q}The static $q_{D}=q\,(T=T_{D}^-)$ and
dynamic $q_{G}=q\,(T=T_{G}^-)$ order parameter as a function of
$1/\log(m)$.  The dotted line is for the static value, the bold line is
for the dynamical one. }
\end{center}
\vspace{8mm}
\end{figure}
Again,  by assuming the jump $q_G$ near the temperature $T_G$ to be small,
one can control the approximation by letting $m=m^*+\varepsilon$, $\varepsilon \ll 1$. Thus,
one has
 \begin{subequations}
\be
q_{G}(m=m^*+\varepsilon) =\frac{3 m^*  g(m^*) (1-2 m^* )  (m^*+4)}{
2(m^*+2)}\, \varepsilon+O[\varepsilon^{2}],
\label{q_G_eps}
\ee
\be
t_{G}(m=m^*+\varepsilon) =\frac{9}{2} \,g(m^*)\, \left(m^{*2}- m^* -5/2\right) \,
\varepsilon ^2 +O[\varepsilon^{3}],
\label{t_G_eps}
\ee
\end{subequations}
Within the approximation used, one finds
\be
\frac{q_G}{q_{D}}=\frac{3}{4}+
\frac{3}{16} g(m^*)  \left(-2 m^{*2}+2 m^*
+5\right) \, \varepsilon^2+
O\left[\varepsilon^3 \right].
\ee
It is worth noting that, to the leading order, exactly the same value for this ratio has
been obtained for the Potts glass in Refs. \cite{desantis, ck, kirkwoly_87}, suggesting a
sort of universality related to the same general structure of the free
energy for both quadrupolar and Potts glass models (see also Ref.
\cite{gold_elder_85}).
The results for $q_{D}$ and $q_G$ as a
function of $1 / \log( m )$ are shown in Fig. \ref{stat_dyn_q}.
The ratio between the two transition temperatures $T_G / T_{D}$ is very
close to one
\be
\frac{T_G}{T_{D}}=1+
\frac{1}{8}\,  g(m^*) \, \left(-2 m^{*2}+2 m^* +5\right)\, \varepsilon^2
+O\left[\varepsilon^3 \right],
\ee
with $T_G$ always bigger than $T_{D}$, as shown in
Fig. \ref{quad_diag_phase}. For $m = 4$ (the smallest integer value of $m$
compatible with a discontinuous transition) one finds that
\be
\begin{split}
q_{D} &=0.056... \quad \quad ; \quad \quad T_{D}=1.342... \\
q_{G} &=0.042...   \quad \quad ; \quad \quad T_{G}=1.343...
\end{split}
\ee
Of course, a naive extension of our result to include large $m$ results
is not possible since we have assumed all the transitions to be at worst
weakly discontinuous, implying the possibility to explore only the range $m$
close to $m^*$.
However, since the mean-field theory of the Potts glass is qualitatively
very similar to that of the quadrupolar glass, one may have an idea of the large
$m$ limit by considering the large $p$ limit in the Potts glass model.  There, the ratio
$q_G/q_{D}$ stays close to $3/4$ for a large range of $p$, though it increases for very
large $p$, approaching unity. The ratio $T_G / T_{D}$ grows very slowly with $p$
\cite{desantis}.
This behaviour is different from the case of the $p$-spin model, where the ratio $q_G/q_{D}$
is not close to $3/4$ even for small values of $p$ and, in the limit $p \to \infty$, it
converges to unity. Moreover, the ratio $T_G / T_{D}$ grows faster with $p$ than in the
Potts problem \cite{desantis}.
\begin{figure}[t]
\begin{center}
\includegraphics*[scale=.35]{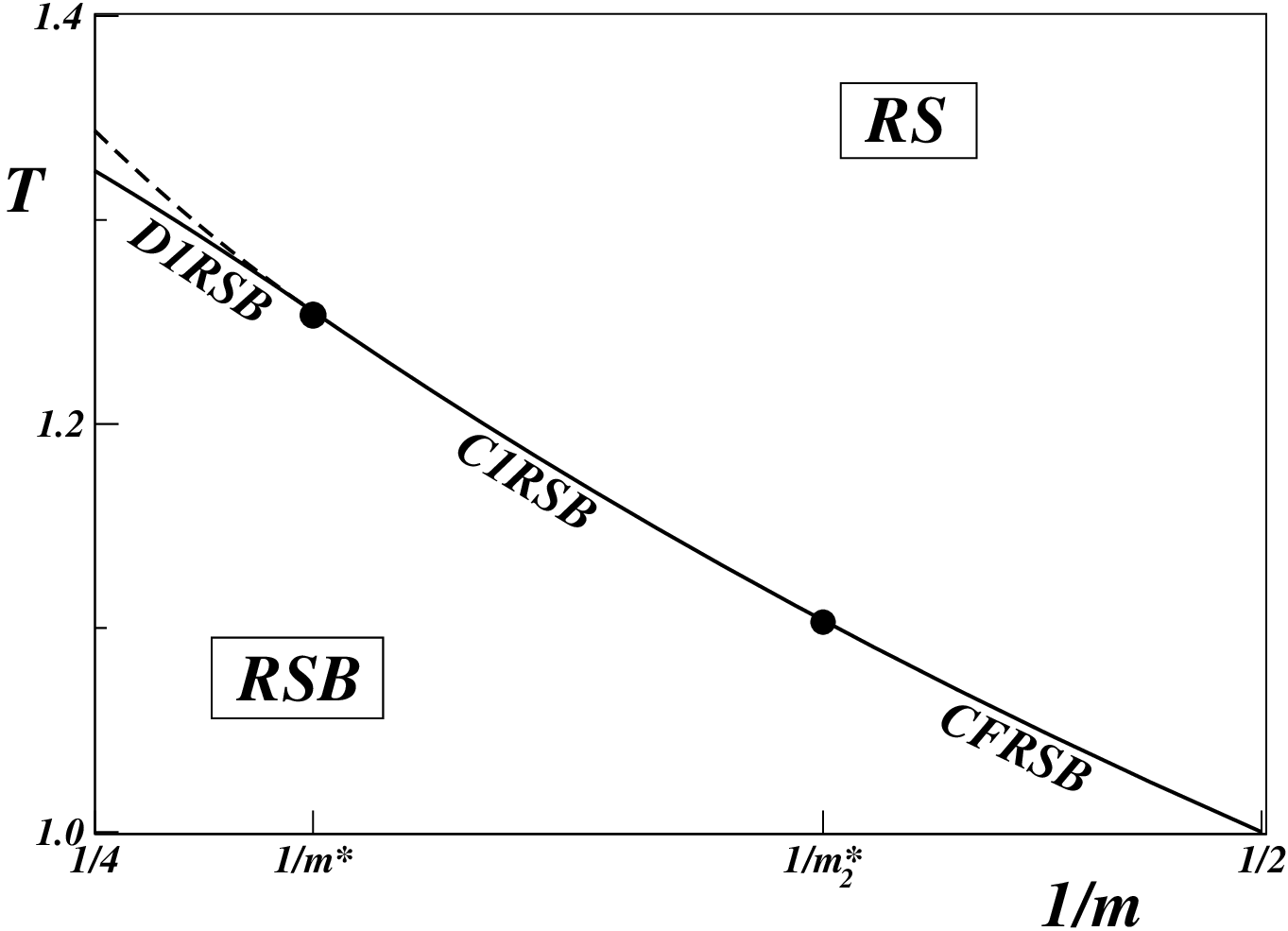}
\caption{\label{quad_diag_phase}
Phase diagram of the quadrupolar glass model in the $(T,1/m)$
plane. $T$ is in units of $k_B/J$. For $2<m<m_2^*$ the transition from the higher
temperature $RS$ phase occurs within the full replica symmetry breaking mechanism.
For $m<m^*$ the thermodynamic and dynamical transitions
coincide. For $m>m^*$ the dynamical transition, denoted by the dotted line, is
higher that the thermodynamic one (solid line). The plot is shown
within the quartic approximation for the free energy expansion in $q$. }
\end{center}
\vspace{8mm}
\end{figure}

A useful and new representation of the phase diagram is obtained
by plotting the phase boundary line in the plane $(T,1/m)$. One
may identify three stable phases, $RS$ paramagnetic and both
$1RSB$ and $FRSB$ glasses. In
Fig. \ref{quad_diag_phase}, the phases are labelled by their
symmetry breaking and the manner of the onset from the paramagnet.
The $1RSB$ transition is continuous between $m_2^* \le m \le m^*$,
whereas it is discontinuous above $m^*$. Note that, at $m=m^*$,
the transition from $RS$ passes continuously from continuous $1RSB$ ($C1RSB$) to
discontinuous $1RSB$ ($D1RSB$) within the one-step $RSB$
phase. The dotted line in the figure corresponds to the $m>m^*$
dynamical transition temperature given in Eq. (\ref{t_G}).
The situation is analogous to that of the Potts glass model which shows a
crossover from the continuous transition to the discontinuous transition as the
number of Potts states increases \cite{gks,ck}.
The $p$$>$$2$-spin Ising and spherical spin glasses also show transitions
from $C1RSB$ to $D1RSB$ as an applied field is reduced but differ from
the present problem in that the critical field makes also a maximum in
the transition temperature, in contrast to the present monotonic
variation with $1/m$. For $m<m_2^*$ the transition is continuous to full replica symmetry
breaking.  A phase line (not shown, but continuous) separates the one-step and full replica
symmetry breaking phases within the $RSB$ region.


\section{Concluding remarks}
\noindent In this paper we have investigated the quadrupolar glass
model in the framework of the replica method. Upon introducing a
simple one-step replica symmetry breaking ansatz, one may find a
stable mean-field theory with a continuous or discontinuous
transition, according to the value of the quadrupole dimension
$m$. The transition is continuous to one-step replica symmetry
breaking in the range of the quadrupolar dimension $2.46<m<3.37$.
For the discontinuous transition ($m>3.37$) there are two
different transition temperatures. We have computed the ratio
$q_G/q_{D}$, where $q_G$ and $q_{D}$ are the order parameters
associated, respectively, with the dynamic and thermodynamic
transition. The ratio between the two transition temperatures
$T_G/T_{D}$ is also computed. Within the approximation used, the
values of these ratios, $q_G/q_{D}=3/4$ and $T_G/T_{D}\simeq 1$
(to leading order), are the same as those found in Refs. \cite{ck,
desantis, kirkwoly_87} for the Potts glass model.

The results we have obtained confirm the general wisdom that the
properties of the quadrupolar glass, with continuous ($m<m^*$) and
discontinuous transitions ($m>m^*$),  are similar to those of the
$p<4$ and $p>4$ Potts glass well studied in the literature
\cite{gks,ck,desantis,kirkthiru2,kirkwoly_87}. Although the
investigation focused on the quadrupolar glass phase, in the wider
$(J_0,J,T)$ space there should exist different types of
ferromagnet, collinear and canted, see e.g. Ref. \cite{gs}.

The full phase diagram should include also another curve which may
be captured by complexity arguments. In analogy with the $p$-spin
spin glass there should be another critical line $T^{compl.}$
associated with the onset of macroscopic complexity
\cite{cavagna99}.

 \section*{Acknowledgements}
We thank P. M. Goldbart for the benefit of stimulating
discussions.  We also thank the EPSRC for
financial support in Oxford under grants GR/R83712/01 and 97304299 and ESF
programme SPHINX for the opportunity to meet to continue this work.


\end{document}